# Microwave Conductivity of Ferroelectric Domains and Domain Walls in Hexagonal Rare-earth Ferrite

Xiaoyu Wu[1], Kai Du[2], Lu Zheng[1], Di Wu[1], Sang-Wook Cheong[2], Keji Lai[1]

[1]Department of Physics, University of Texas at Austin, Austin TX 78712, USA

[2]Rutgers Center for Emergent Materials and Department of Physics and Astronomy, Rutgers University, Piscataway NJ 08854, USA

## Abstract

We report the nanoscale electrical imaging results in hexagonal $Lu_{0.6}Sc_{0.4}FeO_3$ single crystals using conductive atomic force microscopy (C-AFM) and scanning microwave impedance microscopy (MIM). While the dc and ac response of the ferroelectric domains can be explained by the surface band bending, the drastic enhancement of domain wall (DW) ac conductivity is clearly dominated by the dielectric loss due to DW vibration rather than mobile-carrier conduction. Our work provides a unified physical picture to describe the local conductivity of ferroelectric domains and domain walls, which will be important for future incorporation of electrical conduction, structural dynamics, and multiferroicity into high-frequency nano-devices.

Domain walls (DWs) in ferroelectric materials are natural interfaces that can be readily written, erased, and manipulated by external electric fields. Since the discovery of DW conduction in BiFeO$_3$ thin films[1] by conductive atomic-force microscopy (C-AFM), similar phenomena have been reported in a wide range of ferroelectrics including PbZr$_{0.2}$Ti$_{0.8}$O$_3$ (PZT) [2], LiNbO$_3$ [3], BaTiO$_3$ [4], hexagonal manganite h-$R$MnO$_3$ (R = Sc, Y, Ho to Lu) [5-7], and (Ca,Sr)$_3$Ti$_2$O$_7$ [8]. It is now generally accepted that the prominent conductivity difference between domains and DWs is a norm rather than an exception[9]. Consequently, much effort has been made to demonstrate DW-based nano-devices[10], such as nonvolatile memory[11,12], field effect transistors (FETs)[13], reconfigurable channels[14,15], and DW-motion logics[16]. While many functionalities are achieved at zero (dc) or low frequencies, practical devices usually demand much higher operation frequencies. In the giga-Hertz (GHz) range, the dielectric loss due to dipolar relaxation may become significant. In other words, the effective ac conductivity would contain contributions from both mobile carrier conduction and bound charge oscillation. The understanding of DW response in the microwave regime is therefore desirable for the continued research in DW nanoelectronics.

The GHz DW conductivity has been recently studied by scanning microwave impedance microscopy (MIM) in several ferroelectrics[17-19]. In particular, charge-neutral DWs on the (001) surface of h-$R$MnO$_3$, which show vanishingly small electrical conduction at dc[5], exhibit very large ac conductivity at radio frequencies due to the collective DW vibration around its equilibrium position[19]. In this Rapid Communication, we report a combined C-AFM and MIM study on hexagonal ferrite Lu$_{0.6}$Sc$_{0.4}$FeO$_3$, which is isomorphic to h-$R$MnO$_3$ in the crystalline structure[20-23]. While mobile carriers are responsible for the DW dc conduction, the large ac conductivity at 1 GHz is clearly dominated by the dielectric loss due to DW oscillations. By applying a tip bias during the MIM imaging, we observed that the signals on the two types of ferroelectric domains could be described by the surface band bending, whereas the GHz conductivity at the DWs remains largely unchanged. Our work provides a platform to explore the interplay between electrical conduction and structural dynamics in multiferroic DWs and generates new impetus to incorporate nanometer-sized DWs into multifunctional nanoelectronic devices.

Hexagonal rare-earth ferrites h-$R$FeO$_3$ (R = Sc, Y, Ho to Lu) have attracted much interest in the past decade due to the possible coexistence of ferroelectricity and antiferromagnetism at room temperature[20-23]. While only the orthorhombic phase is thermodynamically stable for bulk LuFeO$_3$, it is found that hexagonal LuFeO$_3$ crystals can be stabilized by Sc substitution without the loss of multiferroicity[23]. In this work, bulk Lu$_{0.6}$Sc$_{0.4}$FeO$_3$ (LSFO) crystals (lattice structure shown in Fig. 1a) were grown using the optical floating zone method under 0.8 MPa O$_2$ atmosphere. The crystals were annealed at 1400°C in air for 24 hours and then cooled down to 1200°C with 1°C/h cooling rates, followed by the final annealing at 1000°C under 20 MPa O$_2$ pressure in a high-pressure oxygen furnace to remove oxygen vacancies. Analogous to h-$R$MnO$_3$, LSFO crystals after such treatments show weak p-type conduction, presumably due to the interstitial oxygen doping[24]. As shown in Fig. 1b, the two-terminal resistance along the hexagonal c-axis of a LSFO sample (~ 0.5 mm in height and ~ 1 mm$^2$ in area) is around 1 ~ 2 MΩ at high dc bias. Assuming that the contact resistance is insignificant in this regime, one can estimate that the bulk dc conductivity $\sigma_{bulk}^{dc}$ is around 10$^{-4}$ ~ 10$^{-3}$ S/m, which is similar to that of h-$R$MnO$_3$ in earlier works[25].

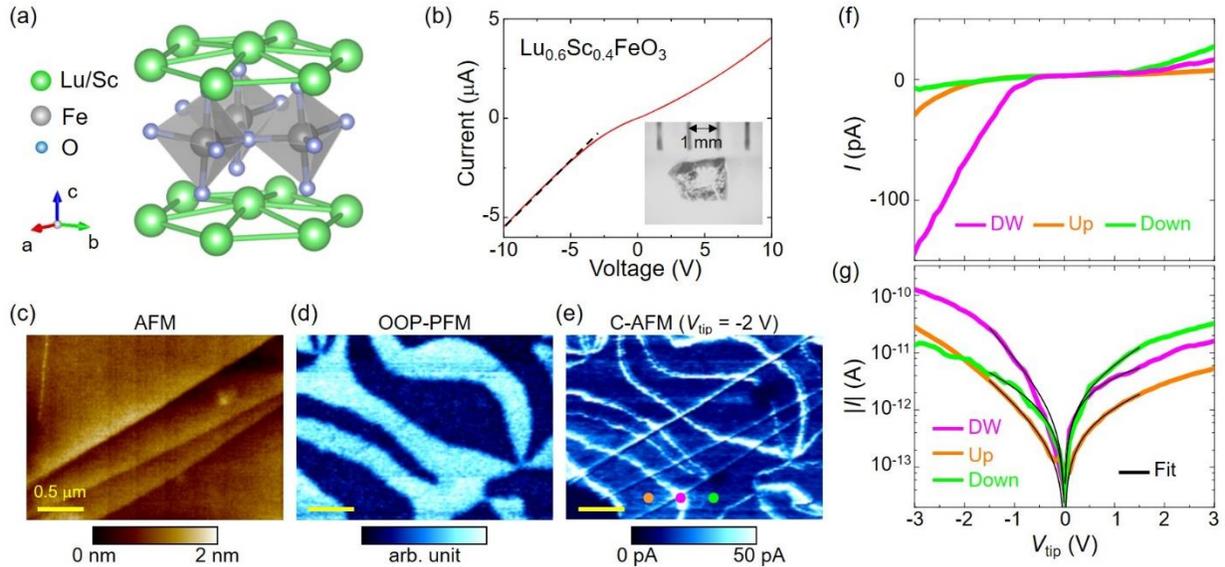

FIG. 1. (a) Crystal structure of hexagonal Lu$_{0.6}$Sc$_{0.4}$FeO$_3$ (LSFO). (b) Two-terminal I-V characteristics on a bulk LSFO single crystal. The dashed line is a linear fit to deduce the resistance. The inset shows a picture of typical LSFO crystals. (c-e) AFM, out-of-plane PFM, and C-AFM images acquired on the same area from the cleaved surface of (001) LSFO. The enhanced C-AFM signals at the step terraces are presumably due to the sudden change of tip-sample contact area during the scanning. The scale bars are 500 nm. (f) Fixed-point I-V curves on up domain (orange), down domain (green), and domain wall (purple) indicated in (e). (g) Semi-log plot of the data in (f). The black lines are exponential fits for |$V_{tip}$| < 1.5 V.

The as-grown LSFO crystals were cleaved to expose the (001) surface for imaging studies. The AFM image in Fig. 1c shows micrometer-scale flat terraces. Cloverleaf-like ferroelectric domain patterns, reminiscent of that observed in h-$R$MnO$_3$, could be seen in the out-of-plane piezo-response force microscopy (OOP-PFM) image (Fig. 1d). Different from h-$R$MnO$_3$ where DWs on the (001) surface are more resistive than the adjacent domains[5], however, the C-AFM image in the same area (Fig. 1e) indicates that the LSFO (001) DWs exhibit enhanced conduction under a tip bias $V_{tip}$ of -2 V. To estimate the local dc conductivity, we measured the fixed-point (labeled in Fig. 1e) $I$-$V$ characteristics on DW and up-polarized/down-polarized domains (hereafter abbreviated as up and down domains, respectively) in Fig. 1f. Similar results can be observed in other locations of the sample. Consistent with an earlier report on HoMnO$_3$[25], the signals on the two domains can be explained by the polarization-modulated rectification at the metal-semiconductor junction. For small bias values $|V_{tip}| < 1.5$ V, the curves can be fitted by the Shockley diode equation $I = I_S[\exp(eV/nk_BT) - 1]$, where $e$ is the electron charge, $k_B$ is the Boltzmann constant, $T$ is the temperature, $I_S$ is the saturation current, and $n$ is the ideality factor[26]. The asymmetric bias-dependent current is presumably due to the tip-sample Schottky barrier. In the high forward bias ($V_{tip} < 0$) regime, the current measured at the DW is ~ 5 times larger than that at the domains. Considering that the typical tip-sample contact diameter of 10 nm is another ~ 5 times larger than the ferroelectric DW width (1~3 nm), we estimate the DW dc conductivity $\sigma_{DW}^{dc}$ to be $10^{-2} \sim 10^{-1}$ S/m, i.e., 1 ~ 2 orders of magnitude higher than $\sigma_{bulk}^{dc}$. We note that several theories have been proposed to explain the enhanced dc conductivity in nominally uncharged DWs, including the band-gap narrowing effect[1,27], accumulation of charged defects[28], and flexoelectric effect[29]. While the exact nature of the DW dc conduction in LSFO is not clear at this point, the level of conductivity enhancement is consistent with other investigations[27-29].

At GHz frequencies, it has been reported that DWs on (001) h-$R$MnO$_3$ exhibit strong dielectric loss due to the periodic vibration around the equilibrium position[19]. Since hexagonal manganites and ferrites share the same lattice structure and origin of ferroelectricity, similar DW dynamics is also expected in the LSFO sample. Fig. 2a shows the MIM images at $f$ = 1 GHz when a dc voltage is applied to the tip through a bias-tee. At zero tip bias, the MIM images are dominated by the pronounced DW signals. As $V_{tip}$ increases from 0 V, contrast between opposite domains emerges, with up domains showing higher MIM signals. The domain contrast reverses sign for a negative tip bias. The evolution of domain signals can be qualitatively described by the band bending at the tip-sample interface[25], as illustrated in Fig. 2b. Here the overall Schottky barrier increases with increasing forward bias ($V_{tip} > 0$) and decreases with increasing absolute reverse bias ($V_{tip} < 0$). Different from non-ferroelectric semiconductors, the polarization-induced surface charge leads to an additional modification to the Schottky barrier height. As a result, in the accumulation regime, the valence band maximum $E_V$ of down domains will reach the Fermi level $E_F$ before the up domains. Conversely, the surface inversion of carrier type when the conduction band minimum $E_C$ meets $E_F$ will occur first at the up domains. At intermediate $V_{tip}$, the semiconductor is in the depletion regime, where $E_F$ is distant from both conduction and valence bands. The polarization-mediated band bending[25] is thus consistent with the domain contrast in the MIM images.

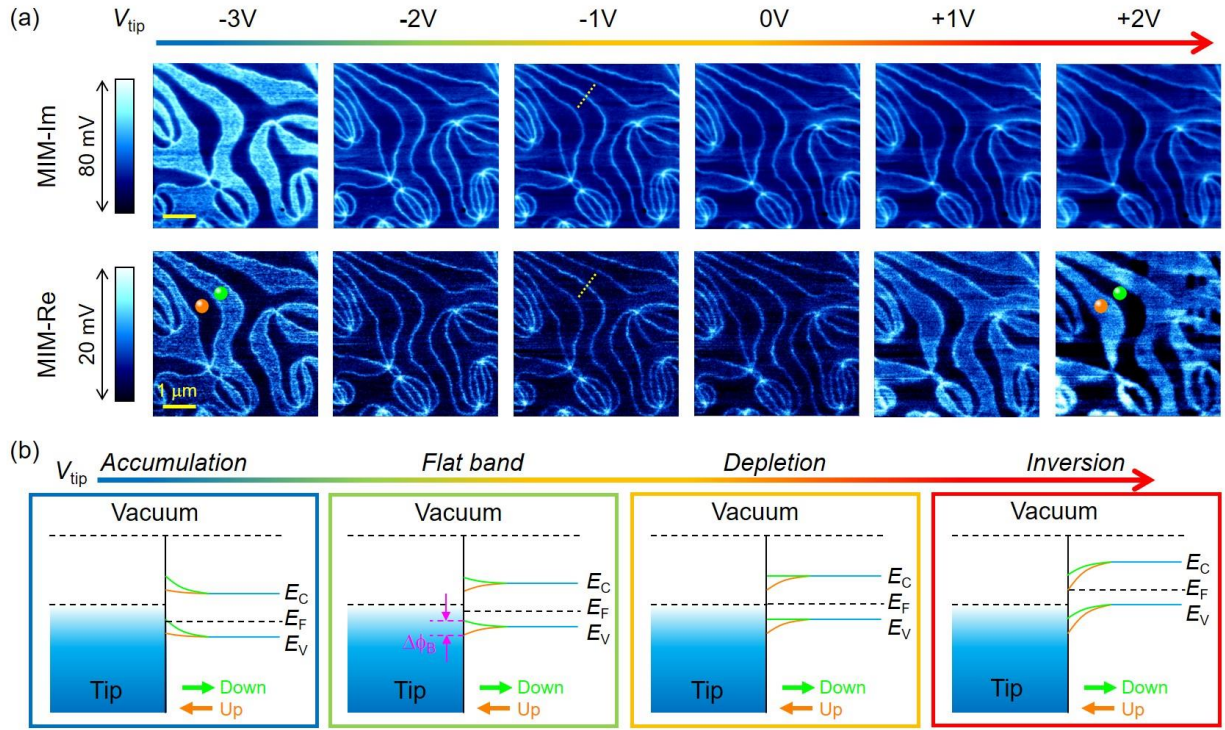

FIG. 2. (a) MIM-Im/Re images on (001) LSFO at different bias voltages. The scale bars are 1 µm. The images were acquired with monotonically increasing $V_{tip}$ from -3 V to +2 V. The up and down domains are marked by orange and green dots, respectively. (b) Schematic diagrams of interfacial band diagrams between the MIM tip and FE domains. $E_C$, $E_V$ and $E_F$ are energy levels of the conduction band, valence band and Fermi energy of semiconductor, respectively. $\Delta\phi_B$ is the difference in Schottky barrier height between the two domains.

To quantify the MIM data, we first analyze the DW signals when both domains are highly resistive, i.e., in the depletion regime. The MIM line profiles across a single DW (indicated in Fig. 2a) are plotted in Fig. 3a. The full width at half maximum (FWHM) of ~ 100 nm is limited by the spatial resolution, which is determined by the diameter of the tip apex that is in close proximity with the sample. The MIM-Im/Re signals are proportional to the real and imaginary parts of the tip-sample admittance, which can be computed by finite-element analysis (FEA)[30]. Here the DW is modeled as a vertical 2-nm-wide slab sandwiched between adjacent insulating domains. The simulated MIM signals as a function of the DW ac conductivity $\sigma_{DW}^{ac}$ are shown in Fig. 3b, from which $\sigma_{DW}^{ac}$ ~ 600 S/m at 1 GHz can be estimated by comparing the measured signals and the FEA results. The fact that $\sigma_{DW}^{ac}$ is about 4 orders of magnitude higher than $\sigma_{DW}^{dc}$ strongly suggests that the DW vibration[19], rather than mobile carrier conduction, is responsible for the energy dissipation at microwave frequencies.

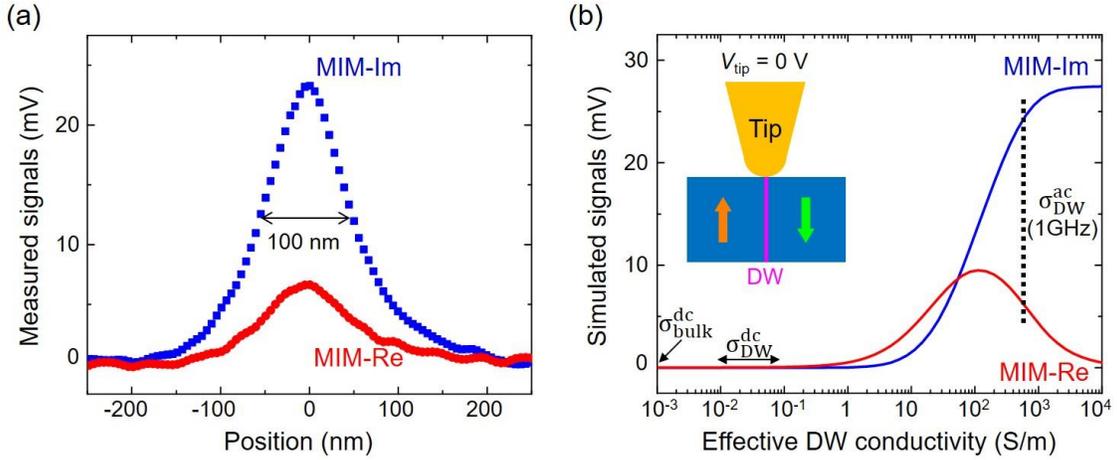

FIG. 3. (a) MIM-Im/Re line profiles across a single DW, labeled as dashed lines in Fig. 2a. The full width at half maximum is ~ 100 nm. (b) Simulated MIM signals as a function of the DW ac conductivity. The signal levels in (a) are consistent with $\sigma_{DW}^{ac}$ ~ 600 S/m, as denoted by the dashed line. $\sigma_{bulk}^{dc}$ and $\sigma_{DW}^{dc}$ are also indicated in the plot. The inset shows the tip-sample configuration for the FEA simulation.

We now turn to the quantitative analysis of bias-dependent MIM images. As discussed before, the surface conductivity is modified by the tip bias due to the band bending at the tip-sample interface. In principle, the spatial distribution of conductivity underneath the biased tip can be numerically computed by self-consistent Schrodinger-Poisson equations[31]. Such an approach, however, requires extensive knowledge on the band parameters, carrier mobility, and the exact tip-sample contact conditions, which are difficult to evaluate from our data. Since the dimension of the tip-sample contact area is much smaller than that of the space charge region, we approximate the tip-induced surface effect by a semi-spherical region (radius $r_{surf} = 100$ nm) with a uniform conductivity $\sigma_{surf}$. The MIM response as a function of $\sigma_{surf}$ is included in Appendix A. Moreover, using a simple dielectric gap model[32,33], one can estimate the difference in the Schottky barrier ($\Delta\phi_B$) between the two domains to be 0.1 ~ 0.2 eV. As a result, for the same $V_{tip}$, the surface conductivity differs by a factor of 100 ~ 1000 when the tip scans across the DW. Since the MIM signals saturate for $\sigma_{surf} < 10^{-2}$ S/m (Appendix A), we further assume that the less conductive domain for a given $V_{tip}$ is at the insulating limit, as schematically illustrated in Fig. 4a. The bias-dependent MIM data across the DW (indicated in Fig. 2a) are plotted in Fig. 4b and 4c. The averaged MIM signals on the DW and up/down domains over the entire image are shown in Figs. 4d and 4e, using the less conductive domain as a reference. Using the tip-sample configuration in Fig. 4a, we can simulate the line profiles (overlaid in Fig. 4b and 4c) and the results are in good agreement with the measured data. Fig. 4f summarizes the calculated $\sigma_{surf}$ and $\sigma_{DW}^{ac}$ from the simulation. Again, while the domain signals can be described by the Schottky band bending, the large $\sigma_{DW}^{ac}$ with virtually no bias dependence signifies the strong dynamic response of ferroelectric DWs at GHz frequencies.

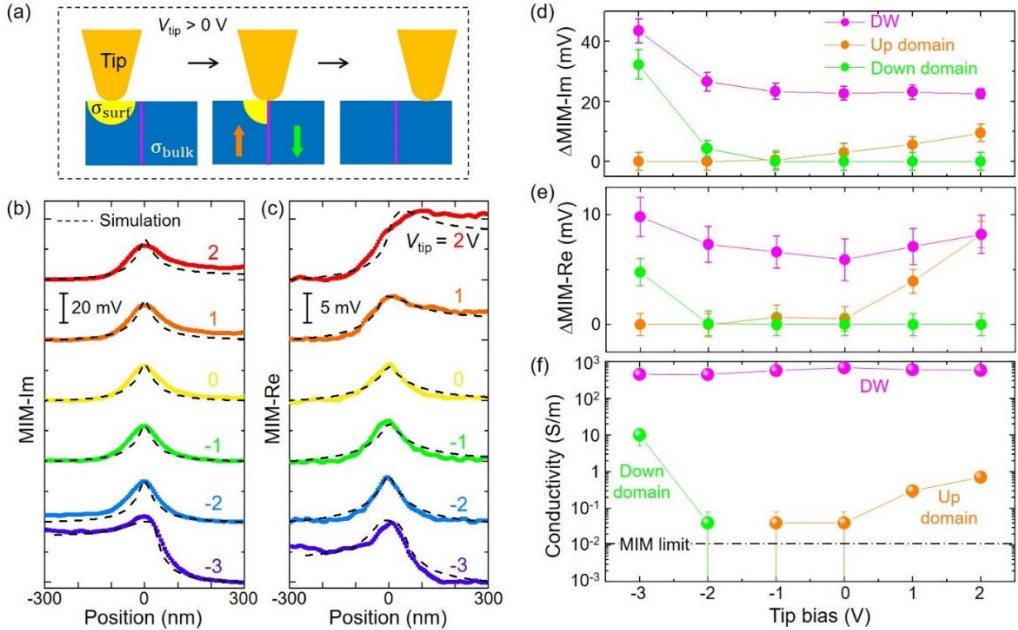

FIG. 4. (a) Schematic of the tip-sample configuration when a positively biased tip scans across the DW. The yellow hemisphere represents the region with enhanced surface conductivity $\sigma_{surf}$. (b) MIM-Im and (c) MIM-Re line profiles across a single DW at different tip biases. The dashed lines show simulated MIM signals to fit the experimental data. (d) Averaged MIM-Im and (e) MIM-Re signals in Fig. 2a, using the less conductive domain as the reference. (f) Surface conductivity of domains and ac conductivity of DWs as a function of $V_{tip}$. The dash-dotted line indicates the MIM sensitivity floor when measuring $\sigma_{surf}$.

Finally, the implications of our results are briefly discussed. In conventional C-AFM measurements, a high $V_{tip}$ is usually required to inject a current across the Schottky barrier. The current then enters the surface accumulation or inversion region and finds its way to the counter electrode through an intricate matrix of domains and DWs. The measured conductance is largely limited by the bulk semiconductor, making it formidable to quantify the local $\sigma^{dc}$. The MIM, on the other hand, probes the local ac impedance by an oscillating GHz voltage ($V_{tip}^{ac} < 0.1$ V) through the near-field interaction, which decays rapidly away from the tip[19,30]. As a result, it is straightforward to interpret the MIM data as an averaged response over a spatial extent determined by the tip diameter. For the ferroelectric domains, the extracted surface conductivity (Fig. 4f) can be satisfactorily explained by the band-bending picture[25] (Fig. 2b), which is not surprising since electrical conduction due to mobile carriers does not differ much between dc and $f = 1$ GHz. In contrast, the measured $\sigma_{DW}^{ac}$ at 1 GHz is nearly bias-independent and the value is ~ $10^4$ times greater than $\sigma_{DW}^{dc}$, indicative of the predominance of dielectric loss due to DW vibration at microwave frequencies[19]. Our results thus provide a unified physical picture to analyze nanoscale dc and ac response of ferroelectric domains and domain walls, which will be invaluable for future DW nanoelectronics operating in the microwave regime.

In summary, we performed electrical mapping on (001) hexagonal $Lu_{0.6}Sc_{0.4}FeO_3$ single crystals at dc and GHz frequencies by a combination of C-AFM and MIM techniques. The dc conductivity of the DWs is moderately enhanced over that of the domains owing to the excess mobile carriers. MIM studies demonstrate that the microwave response of DWs is dominated by their vibrational dynamics, resulting in a bias-independent effective ac conductivity higher than the dc value by a factor of ~ $10^4$. As h-$Lu_{0.6}Sc_{0.4}FeO_3$ is a room-temperature multiferroic material[23,34], our results shed new lights on the interplay among electrical conduction, structural dynamics, ferroelectricity, and magnetism in small band-gap multiferroics.

**ACKNOWLEDGMENTS**

The MIM work (X.W., L.Z., D.W., K.L.) was supported by NSF Award DMR-1707372. The work at Rutgers (K.D., S.-W. C.) was supported by the Gordon and Betty Moore Foundation's EPiQS Initiative through Grant GBMF4413 to the Rutgers Center for Emergent Materials. The authors thank W. Wu for helpful discussions.

**APPENDIX A: Finite-element analysis of the MIM results**

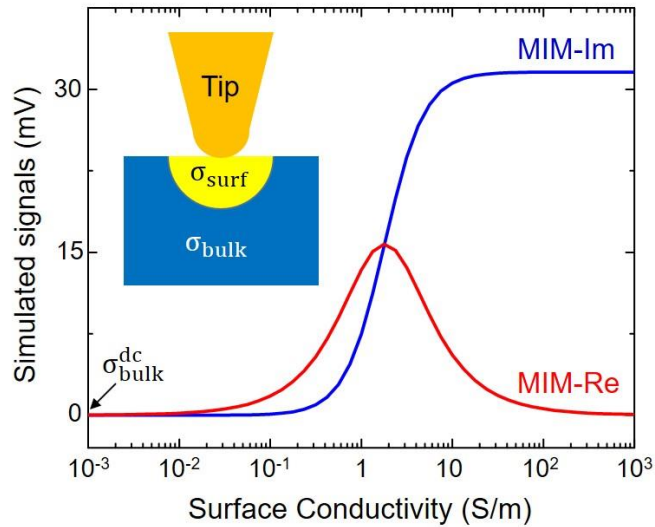

FIG. A1. Simulated MIM signals as a function of the surface conductivity of the domains. The bulk dc conductivity is also indicated in the plot. The inset shows a side view of the tip-sample geometry for the FEA simulation.

Finite-element analysis (FEA) of the MIM data[30] was performed by the commercial software COMSOL 4.4. The tip diameter of 100 nm is the same as the full width at half maximum of the MIM line profile across the DW, which is modeled as a 2-nm slab sandwiched between up and down domains[19]. To analyze the bias-dependent images, we assume that $V_{tip}$ induces a semi-spherical region ($r_{surf}$ = 100 nm) underneath the tip with a uniform conductivity $\sigma_{surf}$. Fig. A1 shows the FEA results as a function of $\sigma_{surf}$ using the 2D axisymmetric model. The MIM signals saturate for $\sigma_{surf}$ below $10^{-2}$ S/m so that the bulk domains ($\sigma_{bulk}$ ~ $10^{-3}$ S/m) can be regarded as in the insulating limit. Note that the tip-sample configurations involving the DW (Fig. 3b inset and Fig. 4a) are no longer axisymmetric around the tip. The full 3D FEA is thus needed to generate the simulation results in Fig. 3b and Fig. 4b. The simulated curves in Fig. 3b are also different from that in Fig. A1.

**References:**

1. J. Seidel, L. W. Martin, Q. He, Q. Zhan, Y. H. Chu, A. Rother, M. E. Hawkridge, P. Maksymovych, P. Yu, M. Gajek, N. Balke, S. V. Kalinin, S. Gemming, F. Wang, G. Catalan, J. F. Scott, N. A. Spaldin, J. Orenstein, and R. Ramesh, "Conduction at domain walls in oxide multiferroics", *Nature Mater*. **8**, 229 (2009).

2. J. Guyonnet, I. Gaponenko, S. Gariglio, and P. Paruch, "Conduction at Domain Walls in Insulating Pb(Zr$_{0.2}$Ti$_{0.8}$)O$_3$ Thin Films", *Adv. Mater*. **23**, 5377 (2011).

3. M. Schröder, A. Haußmann, A. Thiessen, E. Soergel, T. Woike, and L. M. Eng, "Conducting Domain Walls in Lithium Niobate Single Crystals", *Adv. Funct. Mater*. **22**, 3936 (2012).

4. T. Sluka, A. K. Tagantsev, P. Bednyakov, and N. Setter, "Free-electron gas at charged domain walls in insulating BaTiO$_3$", *Nature Commun*. **4**, 1808 (2013).

5. T. Choi, Y. Horibe, H. T. Yi, Y. J. Choi, W. Wu, and S.-W. Cheong, "Insulating interlocked ferroelectric and structural antiphase domain walls in multiferroic YMnO$_3$", *Nature Mater*. **9**, 253 (2010).

6. D. Meier, J. Seidel, A. Cano, K. Delaney, Y. Kumagai, M. Mostovoy, N. A. Spaldin, R. Ramesh, and M. Fiebig, "Anisotropic conductance at improper ferroelectric domain walls", *Nature Mater*. **11**, 284 (2012).

7. W. Wu, Y. Horibe, N. Lee, S. W. Cheong and J. R. Guest, "Conduction of Topologically Protected Charged Ferroelectric Domain Walls", *Phys. Rev. Lett*. 108, 077203 (2012).

8. Y. S. Oh, X. Luo, F.-T. Huang, Y. Wang, and S.-W. Cheong, "Experimental demonstration of hybrid improper ferroelectricity and the presence of abundant charged walls in (Ca,Sr)$_3$Ti$_2$O$_7$ crystals", *Nature Mater*. **14**, 407 (2015).

9. R. K. Vasudevan, W. Wu, J. R. Guest, A. P. Baddorf, A. N. Morozovska, E. A. Eliseev, N. Balke, V. Nagarajan, P. Maksymovych, and S. V. Kalinin, "Domain Wall Conduction and Polarization-Mediated Transport in Ferroelectrics", *Adv. Funct. Mater*. **23**, 2592 (2013).

10. G. Catalan, J. Seidel, R. Ramesh, and J. F. Scott, "Domain wall nanoelectronics", *Rev. Mod. Phys*. **84**, 119 (2012).

11. P. Sharma, Q. Zhang, D. Sando, C. H. Lei, Y. Liu, J. Li, V. Nagarajan, and J. Seidel, "Nonvolatile ferroelectric domain wall memory", *Sci. Adv*. **3**, e1700512 (2017).

12. J. Jiang, Z. L. Bai, Z. H. Chen, L. He, D. W. Zhang, Q. H. Zhang, J. A. Shi, M. H. Park, J. F. Scott, C. S. Hwang, and A. Q. Jiang, "Temporary formation of highly conducting domain walls for non-destructive read-out of ferroelectric domain-wall resistance switching memories", *Nature Mater*. **17**, 49 (2017).

13. J. A. Mundy, J. Schaab, Y. Kumagai, A. Cano, M. Stengel, I. P. Krug, D. M. Gottlob, H. Doğanay, M. E. Holtz, R. Held, Z. Yan, E. Bourret, C. M. Schneider, D. G. Schlom, D. A. Muller, R. Ramesh, N. A. Spaldin, and D. Meier, "Functional electronic inversion layers at ferroelectric domain walls", *Nature Mater*. **16**, 622 (2017).

14. I. Stolichnov, L. Feigl, L. J. McGilly, T. Sluka, X.-K. Wei, E. Colla, A. Crassous, K. Shapovalov, P. Yudin, A. K. Tagantsev, and N. Setter, "Bent Ferroelectric Domain Walls as Reconfigurable Metallic-Like Channels", *Nano Lett*. **15**, 8049 (2015).

15. J. R. Whyte, R. G. P. McQuaid, P. Sharma, C. Canalias, J. F. Scott, A. Gruverman, and J. M. Gregg, "Ferroelectric Domain Wall Injection", *Adv. Mater*. **26**, 293 (2014).

16. L. J. McGilly, P. Yudin, L. Feigl, A. K. Tagantsev, and N. Setter, "Controlling domain wall motion in ferroelectric thin films", *Nature Nanotech*. **10**, 145 (2015).

17. A. Tselev, P. Yu, Y. Cao, L. R. Dedon, L. W. Martin, S. V. Kalinin, and P. Maksymovych, "Microwave a.c. conductivity of domain walls in ferroelectric thin films", *Nature Commun*. **7**, 11630 (2016).

18. T. T. A. Lummen, J. Leung, A. Kumar, X. Wu, Y. Ren, B. K. Van Leeuwen, R. C. Haislmaier, M. Holt, K. Lai, S. V. Kalinin, and V. Gopalan, "Emergent Low-Symmetry Phases and Large Property Enhancements in Ferroelectric $KNbO_3$ Bulk Crystals", *Adv. Mater*. **29**, 1700530 (2017).

19. X. Wu, U. Petralanda, L. Zheng, Y. Ren, R. Hu, S.-W. Cheong, S. Artyukhin, and K. Lai, "Low-energy structural dynamics of ferroelectric domain walls in hexagonal rare-earth manganites", *Sci. Adv*. **3**, e1602371 (2017).

20. W. Wang, J. Zhao, W. Wang, Z. Gai, N. Balke, M. Chi, H. N. Lee, W. Tian, L. Zhu, X. Cheng, D. J. Keavney, J. Yi, T. Z. Ward, P. C. Snijders, H. M. Christen, W. Wu, J. Shen and X. Xu, "Room-Temperature Multiferroic Hexagonal $LuFeO_3$ Films", *Phys. Rev. Lett*. **110**, 237601 (2013).

21. S. M. Disseler, X. Luo, B. Gao, Y. S. Oh, R. Hu, Y. Wang, D. Quintana, A. Zhang, Q. Huang, J. Lau, R. Paul, J. W. Lynn, S.-W. Cheong, and W. Ratcliff, "Multiferroicity in doped hexagonal $LuFeO_3$", *Phys. Rev. B* **92**, 054435 (2015).

22. S. M. Disseler, J. A. Borchers, C. M. Brooks, J. A. Mundy, J. A. Moyer, D. A. Hillsberry, E. L. Thies, D. A. Tenne, J. Heron, M. E. Holtz, J. D. Clarkson, G. M. Stiehl, P. Schiffer, D. A. Muller, D. G. Schlom, and W. D. Ratcliff, "Magnetic Structure and Ordering of Multiferroic Hexagonal $LuFeO_3$", *Phys. Rev. Lett*. **114**, 217602 (2015).

23. L. Lin, H. M. Zhang, M. F. Liu, S. Shen, S. Zhou, D. Li, X. Wang, Z. B. Yan, Z. D. Zhang, J. Zhao, S. Dong, and J. M. Liu, "Hexagonal phase stabilization and magnetic orders of multiferroic $Lu_{1-x}Sc_xFeO_3$", *Phys. Rev. B* **93**, 075146 (2016).

24. S. H. Skjærvø, E. T. Wefring, S. K. Nesdal, N. H. Gaukås, G. H. Olsen, J. Glaum, T. Tybell, and S. M. Selbach, "Interstitial oxygen as a source of p-type conductivity in hexagonal manganites", *Nature Commun*. **7**, 13745 (2016).

25. W. Wu, J. R. Guest, Y. Horibe, S. Park, T. Choi, S. W. Cheong, and M. Bode, "Polarization-Modulated Rectification at Ferroelectric Surfaces", *Phys. Rev. Lett*. **104**, 217601 (2010).

26. S. M. Sze, *Physics of Semiconductor Devices* (John Wiley and Sons, Inc., New York, 1981).

27. A. Lubk, S. Gemming, and N. A. Spaldin, "First-principles study of ferroelectric domain walls in multiferroic bismuth ferrite", *Phys. Rev. B* **80**, 104110 (2009).

28. T. Rojac, A. Bencan, G. Drazic, N. Sakamoto, H. Ursic, B. Jancar, G. Tavcar, M. Makarovic, J. Walker, B. Malic, and D. Damjanovic, "Domain-wall conduction in ferroelectric $BiFeO_3$ controlled by accumulation of charged defects", *Nature Mater*. **16**, 322 (2016).

29. E. A. Eliseev, A. N. Morozovska, G. S. Svechnikov, P. Maksymovych, and S. V. Kalinin, "Domain wall conduction in multiaxial ferroelectrics", *Phys. Rev. B* **85**, 045312 (2012).

30. K. Lai, W. Kundhikanjana, M. Kelly, and Z. X. Shen, "Modeling and characterization of a cantilever-based near-field scanning microwave impedance microscope", *Rev. Sci. Instrum*. **79**, 063703 (2008).

31. I-H. Tan, G. L. Snider, L. D. Chang, and E. L. Hu, "A self-consistent solution of Schrödinger-Poisson equations using a nonuniform mesh", *J. Appl. Phys*. **68**, 4071 (1990).

32. L. Pintilie and M. Alexe, "Metal-ferroelectric-metal heterostructures with Schottky contacts. I. Influence of the ferroelectric properties", *J. Appl. Phys*. **98**, 124103 (2005).

33. L. Pintilie, I. Boerasu, M. J. M. Gomes, T. Zhao, R. Ramesh, and M. Alexe, "Metal-ferroelectric-metal structures with Schottky contacts. II. Analysis of the experimental current-voltage and capacitance-voltage characteristics of $Pb(Zr,Ti)O_3$ thin films", *J. Appl. Phys*. **98**, 124104 (2005).

34. K. Du, B. Gao, Y. Wang, X. Xu, J. Kim, R. Hu, F.-T. Huang & S.-W. Cheong, "Vortex ferroelectric domains, large-loop weak ferromagnetic domains, and their decoupling in hexagonal $(Lu, Sc)FeO_3$", *npj Quantum Mater*. **3**, 33 (2018).